\documentclass[11pt]{article}
\usepackage{graphicx}
\usepackage{dcolumn}     
\usepackage{bm}

\newcommand{\BABARPubYear}    {05}
\newcommand{\BABARConfNumber} {131}
\newcommand{\SLACPubNumber} {11606}

\usepackage{relsize}
\def\babar{\mbox{\slshape B\kern-0.1em{\smaller A}\kern-0.1em
    B\kern-0.1em{\smaller A\kern-0.2em R}}}

\def\en         {\ensuremath{e^-}\xspace}   
\def\ep         {\ensuremath{e^+}\xspace}

\def\mup        {\ensuremath{\mu^+}\xspace}
\def\mun        {\ensuremath{\mu^-}\xspace} 
\def\mumu       {\ensuremath{\mu^+\mu^-}\xspace}

\def\taup       {\ensuremath{\tau^+}\xspace}
\def\taum       {\ensuremath{\tau^-}\xspace}

\def\ellell     {\ensuremath{\ell^+ \ell^-}\xspace}

\def\nub        {\ensuremath{\overline{\nu}}\xspace}
\def\nunub      {\ensuremath{\nu{\overline{\nu}}}\xspace}
\def\nub        {\ensuremath{\overline{\nu}}\xspace}
\def\nunub      {\ensuremath{\nu{\overline{\nu}}}\xspace}
\def\nue        {\ensuremath{\nu_e}\xspace}
\def\nueb       {\ensuremath{\nub_e}\xspace}

\def\num        {\ensuremath{\nu_\mu}\xspace}
\def\numb       {\ensuremath{\nub_\mu}\xspace}

\def\nut        {\ensuremath{\nu_\tau}\xspace}
\def\nutb       {\ensuremath{\nub_\tau}\xspace}

\def\nul        {\ensuremath{\nu_\ell}\xspace}
\def\nulb       {\ensuremath{\nub_\ell}\xspace}

\def\g     {\ensuremath{\gamma}\xspace}
\def\gaga  {\ensuremath{\gamma\gamma}\xspace}  
\def\ggstar{\ensuremath{\gamma\gamma^*}\xspace}

\def\piz   {\ensuremath{\pi^0}\xspace}

\def\ppz   {\ensuremath{\pi^0\pi^0}\xspace}
\def\pip   {\ensuremath{\pi^+}\xspace}
\def\pim   {\ensuremath{\pi^-}\xspace}
\def\pipi  {\ensuremath{\pi^+\pi^-}\xspace}

\def\pimp  {\ensuremath{\pi^\mp}\xspace}

\def\kaon  {\ensuremath{K}\xspace}
\def\Kbar  {\kern 0.2em\overline{\kern -0.2em K}{}\xspace}

\def\Kz    {\ensuremath{K^0}\xspace}
\def\Kzb   {\ensuremath{\Kbar^0}\xspace}
\def\KzKzb {\ensuremath{\Kz \kern -0.16em \Kzb}\xspace}
\def\Kp    {\ensuremath{K^+}\xspace}
\def\Km    {\ensuremath{K^-}\xspace}
\def\Kpm   {\ensuremath{K^\pm}\xspace}

\def\KpKm  {\ensuremath{\Kp \kern -0.16em \Km}\xspace}
\def\KS    {\ensuremath{K^0_{\scriptscriptstyle S}}\xspace} 
\def\KL    {\ensuremath{K^0_{\scriptscriptstyle L}}\xspace} 
\def\Kstarz  {\ensuremath{K^{*0}}\xspace}

\def\Kstar   {\ensuremath{K^*}\xspace}

\def\Kstarp  {\ensuremath{K^{*+}}\xspace}

\def\Dbar    {\kern 0.2em\overline{\kern -0.2em D}{}\xspace}

\def\Dz      {\ensuremath{D^0}\xspace}
\def\Dzb     {\ensuremath{\Dbar^0}\xspace}
\def\DzDzb   {\ensuremath{\Dz {\kern -0.16em \Dzb}}\xspace}
\def\Dp      {\ensuremath{D^+}\xspace}
\def\Dm      {\ensuremath{D^-}\xspace}

\def\DpDm    {\ensuremath{\Dp {\kern -0.16em \Dm}}\xspace}
\def\Dstar   {\ensuremath{D^*}\xspace}

\def\Dstarz  {\ensuremath{D^{*0}}\xspace}

\def\Dstarp  {\ensuremath{D^{*+}}\xspace}
\def\Dstarm  {\ensuremath{D^{*-}}\xspace}

\def\B       {\ensuremath{B}\xspace}
\def\Bbar    {\kern 0.18em\overline{\kern -0.18em B}{}\xspace}
\def\Bb      {\ensuremath{\Bbar}\xspace}
\def\BB      {\ensuremath{\B {\kern -0.16em \Bb}}\xspace}
\def\Bz      {\ensuremath{B^0}\xspace}
\def\Bzb     {\ensuremath{\Bbar^0}\xspace}
\def\BzBzb   {\ensuremath{\Bz {\kern -0.16em \Bzb}}\xspace}

\def\Bu      {\ensuremath{B^+}\xspace}
\def\Bub     {\ensuremath{B^-}\xspace}

\def\Bpm     {\ensuremath{B^\pm}\xspace}

\def\BpBm    {\ensuremath{\Bu {\kern -0.16em \Bub}}\xspace}

\def\BorBbar    {\kern 0.18em\optbar{\kern -0.18em B}{}\xspace}
\def\DorDbar    {\kern 0.18em\optbar{\kern -0.18em D}{}\xspace}
\def\KorKbar    {\kern 0.18em\optbar{\kern -0.18em K}{}\xspace}

\def\jpsi     {\ensuremath{{J\mskip -3mu/\mskip -2mu\psi\mskip 2mu}}\xspace}

\mathchardef\Upsilon="7107
\def\Y#1S{\ensuremath{\Upsilon{(#1S)}}\xspace}

\def\FourS {\Y4S}

\mathchardef\Deltares="7101
\mathchardef\Xi="7104
\mathchardef\Lambda="7103
\mathchardef\Sigma="7106
\mathchardef\Omega="710A

\def\Deltabar{\kern 0.25em\overline{\kern -0.25em \Deltares}{}\xspace}
\def\Lbar{\kern 0.2em\overline{\kern -0.2em\Lambda\kern 0.05em}\kern-0.05em{}\xspace}
\def\Sigbar{\kern 0.2em\overline{\kern -0.2em \Sigma}{}\xspace}
\def\Xibar{\kern 0.2em\overline{\kern -0.2em \Xi}{}\xspace}
\def\Obar{\kern 0.2em\overline{\kern -0.2em \Omega}{}\xspace}
\def\Nbar{\kern 0.2em\overline{\kern -0.2em N}{}\xspace}
\def\Xb{\kern 0.2em\overline{\kern -0.2em X}{}\xspace}

\def\X {\ensuremath{X}\xspace}

\def\BR         {{\ensuremath{\cal B}\xspace}}

\def\bpsikl     {\ensuremath{\Bz \to \jpsi \KL}\xspace}

\def\Bzbtox     {\ensuremath{\Bzb \to \X}\xspace}

\def\mus  {\ensuremath{\rm \,\mus}\xspace}

\def\mus        {\ensuremath{\,\mu{\rm s}}\xspace}    

\def\degrees{\ensuremath{^{\circ}}\xspace}

\def\to                 {\ensuremath{\rightarrow}\xspace}

\def\pep2{PEP-II}
\def\BF{$B$ Factory}

\def\gsim{{~\raise.15em\hbox{$>$}\kern-.85em
          \lower.35em\hbox{$\sim$}~}\xspace}
\def\lsim{{~\raise.15em\hbox{$<$}\kern-.85em
          \lower.35em\hbox{$\sim$}~}\xspace}

\def\epstag  {\ensuremath{\varepsilon_{\rm tag}}\xspace}

\def\jetset74   {\mbox{\tt Jetset \hspace{-0.5em}7.\hspace{-0.2em}4}\xspace}

\usepackage{relsize}

\def\fzz{\ensuremath{f_{00}}}

\input babarsym

\long\def\inst#1{\par\nobreak\kern 4pt\nobreak
    {\it #1}\par\vskip 10pt plus 3pt minus 3pt}

\RequirePackage{xspace}

\usepackage{relsize}
\def\babar{\mbox{\slshape B\kern-0.1em{\smaller A}\kern-0.1em
    B\kern-0.1em{\smaller A\kern-0.2em R}}}

\def\en         {\ensuremath{e^-}\xspace}   
\def\ep         {\ensuremath{e^+}\xspace}

\def\mup        {\ensuremath{\mu^+}\xspace}
\def\mun        {\ensuremath{\mu^-}\xspace} 
\def\mumu       {\ensuremath{\mu^+\mu^-}\xspace}

\def\taup       {\ensuremath{\tau^+}\xspace}
\def\taum       {\ensuremath{\tau^-}\xspace}

\def\ellell     {\ensuremath{\ell^+ \ell^-}\xspace}

\def\nub        {\ensuremath{\overline{\nu}}\xspace}
\def\nunub      {\ensuremath{\nu{\overline{\nu}}}\xspace}
\def\nub        {\ensuremath{\overline{\nu}}\xspace}
\def\nunub      {\ensuremath{\nu{\overline{\nu}}}\xspace}
\def\nue        {\ensuremath{\nu_e}\xspace}
\def\nueb       {\ensuremath{\nub_e}\xspace}

\def\num        {\ensuremath{\nu_\mu}\xspace}
\def\numb       {\ensuremath{\nub_\mu}\xspace}

\def\nut        {\ensuremath{\nu_\tau}\xspace}
\def\nutb       {\ensuremath{\nub_\tau}\xspace}

\def\nul        {\ensuremath{\nu_\ell}\xspace}
\def\nulb       {\ensuremath{\nub_\ell}\xspace}

\def\g     {\ensuremath{\gamma}\xspace}
\def\gaga  {\ensuremath{\gamma\gamma}\xspace}  
\def\ggstar{\ensuremath{\gamma\gamma^*}\xspace}

\def\piz   {\ensuremath{\pi^0}\xspace}

\def\ppz   {\ensuremath{\pi^0\pi^0}\xspace}
\def\pip   {\ensuremath{\pi^+}\xspace}
\def\pim   {\ensuremath{\pi^-}\xspace}
\def\pipi  {\ensuremath{\pi^+\pi^-}\xspace}

\def\pimp  {\ensuremath{\pi^\mp}\xspace}

\def\kaon  {\ensuremath{K}\xspace}
\def\Kbar  {\kern 0.2em\overline{\kern -0.2em K}{}\xspace}

\def\Kz    {\ensuremath{K^0}\xspace}
\def\Kzb   {\ensuremath{\Kbar^0}\xspace}
\def\KzKzb {\ensuremath{\Kz \kern -0.16em \Kzb}\xspace}
\def\Kp    {\ensuremath{K^+}\xspace}
\def\Km    {\ensuremath{K^-}\xspace}
\def\Kpm   {\ensuremath{K^\pm}\xspace}

\def\KpKm  {\ensuremath{\Kp \kern -0.16em \Km}\xspace}
\def\KS    {\ensuremath{K^0_{\scriptscriptstyle S}}\xspace} 
\def\KL    {\ensuremath{K^0_{\scriptscriptstyle L}}\xspace} 
\def\Kstarz  {\ensuremath{K^{*0}}\xspace}

\def\Kstar   {\ensuremath{K^*}\xspace}

\def\Kstarp  {\ensuremath{K^{*+}}\xspace}

\def\Dbar    {\kern 0.2em\overline{\kern -0.2em D}{}\xspace}

\def\Dz      {\ensuremath{D^0}\xspace}
\def\Dzb     {\ensuremath{\Dbar^0}\xspace}
\def\DzDzb   {\ensuremath{\Dz {\kern -0.16em \Dzb}}\xspace}
\def\Dp      {\ensuremath{D^+}\xspace}
\def\Dm      {\ensuremath{D^-}\xspace}

\def\DpDm    {\ensuremath{\Dp {\kern -0.16em \Dm}}\xspace}
\def\Dstar   {\ensuremath{D^*}\xspace}

\def\Dstarz  {\ensuremath{D^{*0}}\xspace}

\def\Dstarp  {\ensuremath{D^{*+}}\xspace}
\def\Dstarm  {\ensuremath{D^{*-}}\xspace}

\def\B       {\ensuremath{B}\xspace}
\def\Bbar    {\kern 0.18em\overline{\kern -0.18em B}{}\xspace}
\def\Bb      {\ensuremath{\Bbar}\xspace}
\def\BB      {\ensuremath{\B {\kern -0.16em \Bb}}\xspace}
\def\Bz      {\ensuremath{B^0}\xspace}
\def\Bzb     {\ensuremath{\Bbar^0}\xspace}
\def\BzBzb   {\ensuremath{\Bz {\kern -0.16em \Bzb}}\xspace}

\def\Bu      {\ensuremath{B^+}\xspace}
\def\Bub     {\ensuremath{B^-}\xspace}

\def\Bpm     {\ensuremath{B^\pm}\xspace}

\def\BpBm    {\ensuremath{\Bu {\kern -0.16em \Bub}}\xspace}

\def\BorBbar    {\kern 0.18em\optbar{\kern -0.18em B}{}\xspace}
\def\DorDbar    {\kern 0.18em\optbar{\kern -0.18em D}{}\xspace}
\def\KorKbar    {\kern 0.18em\optbar{\kern -0.18em K}{}\xspace}

\def\jpsi     {\ensuremath{{J\mskip -3mu/\mskip -2mu\psi\mskip 2mu}}\xspace}

\mathchardef\Upsilon="7107
\def\Y#1S{\ensuremath{\Upsilon{(#1S)}}\xspace}

\def\FourS {\Y4S}

\mathchardef\Deltares="7101
\mathchardef\Xi="7104
\mathchardef\Lambda="7103
\mathchardef\Sigma="7106
\mathchardef\Omega="710A

\def\Deltabar{\kern 0.25em\overline{\kern -0.25em \Deltares}{}\xspace}
\def\Lbar{\kern 0.2em\overline{\kern -0.2em\Lambda\kern 0.05em}\kern-0.05em{}\xspace}
\def\Sigbar{\kern 0.2em\overline{\kern -0.2em \Sigma}{}\xspace}
\def\Xibar{\kern 0.2em\overline{\kern -0.2em \Xi}{}\xspace}
\def\Obar{\kern 0.2em\overline{\kern -0.2em \Omega}{}\xspace}
\def\Nbar{\kern 0.2em\overline{\kern -0.2em N}{}\xspace}
\def\Xb{\kern 0.2em\overline{\kern -0.2em X}{}\xspace}

\def\X {\ensuremath{X}\xspace}

\def\BR         {{\ensuremath{\cal B}\xspace}}

\def\bpsikl     {\ensuremath{\Bz \to \jpsi \KL}\xspace}

\def\Bzbtox     {\ensuremath{\Bzb \to \X}\xspace}

\def\mus  {\ensuremath{\rm \,\mus}\xspace}

\def\mus        {\ensuremath{\,\mu{\rm s}}\xspace}    

\def\degrees{\ensuremath{^{\circ}}\xspace}

\def\to                 {\ensuremath{\rightarrow}\xspace}

\def\pep2{PEP-II}
\def\BF{$B$ Factory}

\def\gsim{{~\raise.15em\hbox{$>$}\kern-.85em
          \lower.35em\hbox{$\sim$}~}\xspace}
\def\lsim{{~\raise.15em\hbox{$<$}\kern-.85em
          \lower.35em\hbox{$\sim$}~}\xspace}

\def\epstag  {\ensuremath{\varepsilon_{\rm tag}}\xspace}

\def\jetset74   {\mbox{\tt Jetset \hspace{-0.5em}7.\hspace{-0.2em}4}\xspace}

\usepackage{relsize}

\def\fzz{\ensuremath{f_{00}}}

\begin{document}
{\pagestyle{empty}

\begin{flushright}
SLAC-PUB-\SLACPubNumber \\
\babar-CONF-\BABARPubYear/\BABARConfNumber \\
January 2006 \\
\end{flushright}

\par\vskip 4cm


\begin{center}
\Large \bf \boldmath Measurements of $|V_{cb}|$ and $|V_{ub}|$ at \babar\ 
\end{center}
\bigskip

\begin{center}
\large Romulus Godang (representing the \babar\ Collaboration)    \\
\vspace{0.35cm}
Department of Physics and Astronomy, University of Mississippi-Oxford,\\
University, MS 38677 \ USA
\mbox{}
\vspace{1cm}

\end{center}
\bigskip \bigskip

\begin{center}
\large {\bf Abstract}
\end{center}
 
We report on new measurements of the Cabibbo-Kobayashi-Maskawa matrix elements 
$|V_{cb}|$ and $|V_{ub}|$ with inclusive and exclusive semileptonic $B$ decays, 
highlighting the recent precision measurements with the \babar\ detector at 
the PEP-II asymmetric-energy $B$ Factory at SLAC.

\vfill

\begin{center}

Contributed to the Proceedings of Particles and Nuclei International Conference, PANIC05,\\
October 24 - 28, 2005, Santa Fe, NM USA 

\end{center}

\vspace{1.0cm}
\begin{center}
{\em Stanford Linear Accelerator Center, Stanford University, 
Stanford, CA 94309} \\ \vspace{0.1cm}\hrule\vspace{0.1cm}
Work supported in part by Department of Energy contract DE-AC03-76SF00515.
\end{center}

\newpage

} 

\section{Introduction}

The stringent tests of the Standard Model are currently not limited by the measurements of the
$CP$-Violation parameter sin $2\beta$~\cite{pdg2004} but by the measured 
ratio of the CKM matrix elements $|V_{ub}|/|V_{cb}|$, which determines the length of 
the left side of the Unitary Triangle.

The semileptonic $B$ meson decays to charm and charmless mesons are the primary tool for measuring 
the CKM matrix elements $|V_{cb}|$ and $|V_{ub}|$ because of their simple theoretical description 
at the parton level.  Their relatively large decay rates are proportional to $|V_{cb}|^{2}$ 
or $|V_{ub}|^{2}$, depend on the quark masses $m_{b}$ and $m_{c}$, and allow us to probe the impact
of strong interactions on the bound quark. 

The semileptonic $B$ meson decays can also be used to achieve a precision measurement
of $\fzz \equiv {\cal B}(\Upsilon(4S) \rightarrow \BzBzb)$, which allows to reduce systematic uncertainty 
on many analyses. We measured $\fzz$ using a novel method, which does not require the knowledge 
of $\tau(B^{+})/\tau(B^{0})$ nor rely on isospin symmetry~\cite{godang05}.
The $\fzz$ value is important for measuring absolute $\Upsilon(4S)$ branching
fractions and for measuring $|V_{cb}|$. Experimental studies of the semileptonic $B$ meson decays can be 
broadly categorized into inclusive and exclusive measurements. 

\section{\boldmath $|V_{cb}|$ Measurements}

The CKM matrix element $|V_{cb}|$ can be extracted from the semileptonic $B$ decay rate by correcting 
the strong interaction effects in the parton-level calculations. The semileptonic $B$ decay rate is 
determined from its semileptonic branching fraction and the average $B$ lifetime measurements.
The perturbative and non-perturbative QCD corrections and their uncertainties can be calculated
in the Heavy Quark Expansion (HQE)~\cite{hqe_model}.
In the kinetic-mass scheme, these expansions in $1/m_{b}$ and $\alpha_{s}(m_{b})$ 
have six parameters to order ${\cal{O}}(1/m_{b}^{3})$: the two running kinetic masses of $b$ and $c$ quarks,
$m_{b}(\mu)$ and $m_{c}(\mu)$, and four non-perturbative parameters: $\mu_{\pi}^{2}(\mu)$, 
$\mu_{G}^{2}(\mu)$, $\rho_{D}^{3}(\mu)$, and $\rho_{LS}^{3}(\mu)$, the expectation value of kinetic, chromomagnetic, 
Darwin, and spin-orbit operators, respectively.  All these parameters depend on the scale $\mu$ 
separating short-distance from long-distance QCD effects; the calculations are performed for 
$\mu$ = 1 GeV~\cite{kinetic_scheme}. 

We measured the inclusive $B \to X_{c} \ell \nu$ branching fraction and the six heavy quark parameters
from a fit to the moments of the hadronic mass and electron energy distribution in semileptonic $B$ 
decays, obtaining $|V_{cb}|=(41.4 \pm 0.4 \pm 0.4 \pm 0.6)\times 10^{-3}$,
${\cal B}(B \to X_{c} e \nu) = (10.61 \pm 0.16 \pm 0.06)\%$, 
$m_{c} = (1.18 \pm 0.07 \pm 0.06 \pm 0.02)$ GeV, $m_{b} = (4.61 \pm 0.05 \pm 0.04 \pm 0.02)$ GeV, 
$\mu_{\pi}^{2}=(0.45 \pm 0.04 \pm 0.04 \pm 0.01) $ GeV$^{2}$,
$\mu_{G}^{2}=(0.27 \pm 0.06 \pm 0.03 \pm 0.02)$ GeV$^{2}$, 
$\rho_{D}^{3}=(0.20 \pm 0.02 \pm 0.02 \pm 0.00)$ GeV$^{3}$, and
$\rho_{LS}^{3}=(-0.09 \pm 0.04 \pm 0.07 \pm 0.01)$ GeV$^{3}$, where the errors refer to 
contributions from the experimental errors on the moment measurements and 
the HQE, and other theoretical uncertainties derived from Refs.~\cite{gambino}.
The fit results are fully compatible with independent estimates of $\mu_{G}^{2}=(0.35 \pm 0.07)$ GeV$^{2}$, 
based on the $B^{*}-B$ mass splitting~\cite{gambino}, and of 
$\rho_{LS}^{3}=(-0.15 \pm 0.10)$ GeV$^{3}$, from the heavy-quark sum rules~\cite{bigi}. 
This is to date the most precise measurement of both $|V_{cb}|$ and the $b$-quark mass.

The CKM matrix elements $|V_{cb}|$ can also be extracted from the exclusive semileptonic
$\Bzb \to D^{*+} \ell^{-} \nulb$ as a function of $w$, where $w$ is the product of the four 
velocities of the $\Bzb$ and $D^{*+}$, and corresponds to the relativistic boost $\gamma$ of 
the $D^{*+}$ in the $\Bzb$ rest frame. 
By extrapolating the differential decay rate of $\Bzb \to D^{*+} \ell^{-} \nulb$ to 
the kinematic limit $w \to 1$, we extract the product of $|V_{cb}|$ and the axial form factor 
${\cal A}_{1}(w=1)$.  We combined this measurement with a lattice QCD calculation~\cite{hashimoto} 
of ${\cal A}_{1}(1) = {\cal F}(1) = 0.919\,^{+0.030}_{-0.035}$ to determine 
$|V_{cb}| = (38.7 \pm 0.3 \pm 1.7\,^{+1.5}_{{-1.3}})\times 10^{-3}$~\cite{exc_vcb}, 
where the errors represents the statistical, the systematic, and the uncertainty
in ${\cal A}_{1}(1)$, respectively.

\section{\boldmath $|V_{ub}|$ Measurements}

The inclusive decay rate $B \to X_{u} \ell \nu$ is directly proportional to $|V_{ub}|^{2}$ and 
can be calculated using HQE; however, the extraction of $|V_{ub}|$ is a challenging
task due to a large background from $B \to X_{c} \ell \nu$ decays. 

We have extracted $|V_{ub}|$ using the following techniques: 
a) the measurement of the lepton spectrum above 2.0 GeV/c, i.e. near the kinematic
endpoint for $B \to X_{c} \ell \nu$ decays~\cite{endpoint},
resulting in $|V_{ub}| = (4.44 \pm 0.25\,^{+0.42}_{{-0.38}}\pm 0.22) \times 10^{-3}$; 
b) the measurement of the lepton spectrum combined with $q^{2}$, the momentum transfer 
squared~\cite{neutrinoreco},
resulting in $|V_{ub}| = (3.95 \pm 0.26\,^{+0.58}_{{-0.42}}\pm 0.25) \times 10^{-3}$; 
c) the measurement of the hadron mass distribution below 1.7 GeV/$c^{2}$ and 
$q^{2} > 8$ GeV$^{2}/c^{4}$ in events tagged by the full reconstruction of a hadronic
decay on the second $B$ meson~\cite{hadrecoil}, 
resulting in $|V_{ub}| = (4.65 \pm 0.34\,^{+0.46}_{{-0.38}} \pm 0.23) \times 10^{-3}$.
In all of the above measurements, the errors are due to experimental, shape function, and
theoretical uncertainties.

We have also measured $|V_{ub}|$ in the exclusive semileptonic $B \to \pi \ell \nu$ decays 
based on three different methods: 
a) in untagged events, in which the neutrino momentum is inferred from the missing momentum, i.e. 
the four-momentum is inferred from the difference between the four-momentum
of the colliding-beam particles and sum of the four-momenta of all detected particles in the event.
This measurement is performed separately in five intervals of $q^{2}$ and leads to an independent
measurement of the shape of the form factor. The results agree well with predictions
from lattice QCD and light-cone sum rules~\cite{neutrino_exc}, 
resulting in $|V_{ub}| = (3.82 \pm 0.14 \pm 0.22 \pm 0.11\,^{+0.88}_{-0.52}) \times 10^{-3}$ from 
$B \to \pi \ell \nu$, where the errors are statistical, systematic, the form factor shape, and 
the form factor normalization; 
b) measurement of $B^{0} \to \pi^{-} \ell^{+} \nu$ decays uses events in which
the signal $B$ meson recoils against a $B$ meson that has been reconstructed in
a semileptonic decay $\Bzb \to D^{(*)+} \ell^{-} \nulb$~\cite{semilep_exc}, 
resulting in $|V_{ub}| = (3.3 \pm 0.4 \pm 0.2\,^{+0.8}_{-0.4}) \times 10^{-3}$;
c) measurements of $B^{0} \to \pi^{-} \ell^{+}\nu$ and $B^{+} \to \pi^{0}\ell^{+}\nu$ decays
in $\Upsilon(4S) \to \BB $ events tagged by a fully reconstructed hadronic $B$ decay in three regions of 
$q^{2}$~\cite{fullreco_exc}, resulting in 
$|V_{ub}| = (3.7 \pm 0.3 \pm 0.2\,^{+0.8}_{-0.5})\times 10^{-3}$, 
where the errors of the last two results are statistical, systematic, and the form factor normalization
uncertainties, respectively.

\section{Conclusion}

Precision measurements of the CKM matrix elements $|V_{cb}|$ and $|V_{ub}|$ would significantly
improve the constraints on the Standard Model.  
The current experimental precision of $|V_{cb}|$ is about 2\% and the precision of 
$|V_{ub}|$ is about 8\%, which is dominated by theory uncertainties.

In the next few years, much larger \BB\ data sample will become available
from the $B$ Factories~\cite{bfactory}, PEP-II~\cite{babar_nim}, and KEKB~\cite{belle_nim}. 
We can expect significant improvements in statistics, in our understanding of 
the experimental and theoretical uncertainties, leading to higher 
precision of $|V_{cb}|$ and $|V_{ub}|$.

I would like to thank all members of the \babar\ Collaboration.
This work was supported by the U.S. Department of Energy grant DE-FG02-91ER40622.

\end{document}